\documentclass[review]{elsarticle}
\usepackage{graphicx}
\usepackage{color}
\usepackage{amsmath}
\usepackage{amsfonts}
\usepackage{amssymb}

\usepackage{hyperref}

\journal{Computer Physics Communications}

\begin{document}

\begin{frontmatter}

\title{Energy Probability Distribution Zeros: A Route to Study Phase Transitions}

\author[ufmg]{B. V. Costa}
\ead{bvc@fisica.ufmg.br}

\author[ufmg]{L. A. S. M\'ol\corref{mycorrespondingauthor}}
\cortext[mycorrespondingauthor]{Corresponding author}
\ead{lucasmol@fisica.ufmg.br}

\author[ufop]{J. C. S. Rocha}
\ead{jcsrocha@iceb.ufop.br}

\address[ufmg]{Laborat\'orio de Simula\c c\~ao, Departamento de F\'isica, ICEx \\  Universidade Federal de Minas Gerais, 31720-901 Belo Horizonte, Minas Gerais, Brazil}
\address[ufop]{Departamento de F\'isica, ICEB, Universidade Federal de Ouro Preto, 35400-000 Ouro Preto, Minas Gerais, Brazil}

\begin{abstract}
    In the study of phase transitions a very few models are accessible to exact solution. In the most cases analytical simplifications have to be done or some numerical technique has to be used to get insight about their critical properties. Numerically, the most common approaches are those based in Monte Carlo simulations together finite size scaling analysis. The use of Monte Carlo techniques requires the estimate of quantities like the specific heat or susceptibilities in a wide range of temperature or the construction of the density of states in large intervals of energy. Although many of these techniques are well developed they may be very time consuming when the system size becomes large enough. It should be suitable to have a method that could surpass those difficulties.
    In this work we present an iterative method to study the critical behavior of a system based on the partial knowledge of the complex Fisher zeros set of the partition function. The method is general with advantages over most conventional techniques since it does not need to identify any order parameter \emph{a priori}. The critical temperature and exponents can be obtained with great precision even in the most unamenable cases like the two dimensional $XY$ model. To test the method and to show how it works we applied it to some selected models where the transitions are well known: The 2D Ising, Potts and XY models  and to a homopolymer system. Our choice cover systems with first order, continuous and Berezinskii-Kosterlitz-Thouless transitions as well as the homopolymer that has two pseudo-transitions. The strategy can easily be adapted to any model, classical or quantum, once we are able to build the corresponding energy probability distribution.
\end{abstract}


\end{frontmatter}

\section{Introduction}
    The study of phase transitions is of particular interest due to its practical importance and theoretical richness. Its understanding is the holy grail of thermodynamics with implications in almost all areas of physics. Unfortunately, only a few special cases are accessible to a complete analytical description. In face of that, many approximated techniques devoted to obtain the transition temperature and exponents were developed over the time. Mean field approximations and the renormalization group~\cite{Kadanoff2009,Fisher1998} technique are two of those approaches which have set remarkable advances in the field. Yang and Lee~\cite{Yang-Lee1952} introduced an alternative way to look to phase transitions by using the concept of zeros of the partition function that has shown to be a worthwhile ground to understand phase transitions. Considering the complex fugacity plane, they showed that the partition function zeros density contains all the relevant information about the thermodynamic behavior of system. In particular they showed that in the thermodynamic limit the density of zeros completely determine its critical behavior. In $1964$ Fisher~\cite{Fisher1964} extended their idea to the complex temperature plane (Fisher zeros).

     Beside that, the use of numerical techniques like Molecular Dynamics, Spin Dynamics and Monte Carlo grew dramatically following the accelerated development of fast computers and mainly due to the discovery of new algorithms. Of particular importance are the works of Swendsen and Wang~\cite{Swendsen_Wang,Wolff} developed to reduce critical slowing down in continuous transitions and multicanonical methods (MUCA)~\cite{Muca1,Muca2,Muca3,Muca4,Muca5} developed to overcome the barrier between two coexisting phases in a first order transition. In addition, the use of histograms to obtain the density of states, $g(E)$, have achieved an inconceivable level of sophistication as compared to a few decades ago. These advances have made it possible to obtain quite reliable estimates for thermodynamical quantities in a wide range of energies allowing a step forward in the description of phase transitions. The Broad Histogram~\cite{Murilo1,Murilo2} and the Replica Exchange Wang-Landau~\cite{Wang-Landau,Vogel} are among the state of the art techniques leading those researches. Working together with finite size scaling it is possible to approach the thermodynamic limit in a systematic and efficient way. Nevertheless, the study of phase transitions requires the identification \emph{a priori} of an order parameter. Quantities like the specific heat or susceptibilities have to be evaluated in a wide range of temperature or the density of states has to be built in large intervals of energies.

    In this paper we introduce a new, general and efficient method based on the partial knowledge of the energy probability distribution (EPD) conjugated with an analogue of the Fisher zeros expansion of the partition function that allow us to precisely determine the transition temperature and exponents for a given system. The method has shown to be robust to be successfully extended to any model and has great potential to evoke further rigorous developments in our understanding of phase transitions. In the following we first describe the Fisher's zeros approach. Next we describe how it can be combined with the EPD to obtain the critical behavior of a given system. We demonstrate the method's reliability by comparing our calculations with the exact results for the $2d$ Ising model~\cite{Onsager1944}. Its power is showed by obtaining the critical properties of the $3$ and $6$ states Potts model in two dimensions~\cite{Wu82}, the $2d$ anisotropic Heisenberg model (XY model)~\cite{B,KT,Evertz,PLA_Costa}, and an elastic flexible polymer model \cite{Schnabel2009}. Those applications cover the most relevant types of transitions, since the $6$ states Potts model have a first order transition, $3$ states Potts model has a continuous transition and the $XY$ model has an infinity order transition, also called Berezinskii-Kosterlitz-Thouless (BKT) transition~\cite{B,KT}. Besides that, the method has shown to be reliable to quest for multiple transitions in a richer system, a $13$-monomers polymer. This polymer has two pseudo transitions, solid-globule and globule-coil, which can be identified as first and second order transitions respectively~\cite{Schnabel2011}.

\section{Fisher zeros}
    In the Fisher's zeros approach~\cite{Fisher1964} a phase transition is characterized by a positive real zero of the partition function in the thermodynamic limit. To make it clear we recall that the partition function is given by \cite{Taylor2013}
\begin{equation}
    Z = \sum_{\textbf{E}} g\left( \textbf{E} \right ) e^{-\beta \textbf{E}} = e^{-\beta \varepsilon_0} \sum_{n=1}^{N} g_{\emph{n}} e^{-\beta \emph{n} \varepsilon},
    \label{Equation1}
\end{equation}
\noindent
were $\beta=1/T$ is the inverse temperature, $g\left( \textbf{E} \right )$ is the density of states and we assumed a discrete set of energies $\textbf{E}_n=\varepsilon_0+n\varepsilon$ with $n =0,1,2,\cdots $~\cite{Julio16}. Here, temperature is measured in units of the Boltzmann constant, $k_{B}$. This equation can be rewritten as a polynomial in $ z \equiv e^{- \beta \varepsilon }$
\begin{equation}
   Z =  e^{-\beta \varepsilon_0} \prod_{n=1}^{N} \left( z - z_{n}  \right).
   \label{Equation2}
\end{equation}
\noindent
  For a finite system of volume $L^{d}$ all zeros, $\{ z_j(L) \}_N$, are complex numbers. Because the polynomial coefficients are real, complex roots always appear as conjugated pairs. If the system undergoes a phase transition at $T_c$ the zero $z^\star(T_c(L))$ moves toward the positive real axis as the system size grows. In general if the system undergoes $M$ transitions we expect that the corresponding zeros $\{ z^\star(L) = a^\star(L) + \mathrm{i} b^\star(L) \}_M \in \{ z_j(L) \}_N$, will converge to the infinite volume limit $b^\star(L) \to 0$ as $L \to \infty$ while $\lim_{L \rightarrow \infty} a^\star(L) = a^\star(\infty)$. These zeros characterizing the phase transitions will be named dominant or the leading zeros.
\begin{figure}
    \includegraphics[keepaspectratio,height=0.8\textwidth]{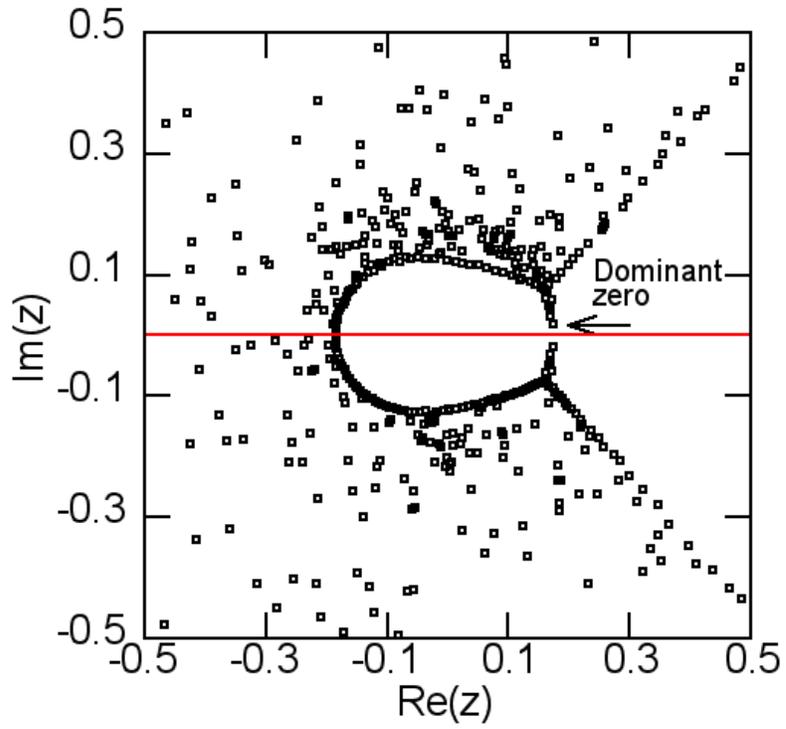}
            \\
      \vspace*{0.75cm}
    \caption{Partial view of the Fisher zeros distribution for the two dimensional Ising model for a $32 \times 32$ lattice. We used the Mathematica's\textsuperscript{\textregistered} solve package to get the zeros. Surprisingly, the expected conjugate pairs of zeros were not found. The reason for this is that the polynomial coefficients span over many orders of magnitude ($10^{0} - 10^{300}$) (See text for details). The arrow marks the \emph{dominant} zero.}
\label{Fig1}
\end{figure}
\noindent
    In Fig.~\ref{Fig1} it is shown a partial view of the distribution of zeros for the $2d$ Ising model in a square lattice of size $L = 32$. The dominant zero is generally the nearest zero to the real positive axis. As discussed above, the zeros must appear as conjugated pairs. However, the Fig.~\ref{Fig1} surprisingly does not show this symmetry. The reason for that will be clear in the following.
    Although $g(E)$ can be calculated exactly in few exceptional cases \cite{exact-dos-ising} and methods like \emph{Wang-Landau}~\cite{Wang-Landau,Vogel} and \emph{Multicanonical}~\cite{Muca1,Muca2,Muca3,Muca4,Muca5}  can give quite good estimates of $g(E)$, the main problem concerning the use of Fisher zeros with those results is that in order to extract the possible transition temperatures we have to build the entire zeros map. This demands one to construct the density of states in the whole energy range. For simpler models and moderate system sizes this is not deterrent. However, as the volume of the system grows, the computer time needed to build $g(E)$ becomes prohibitive. Moreover, for larger system sizes the number of energy states (the polynomial degree) increases dramatically, making the zeros finder task problematic. The density of states spans over many orders of magnitude, making it even worst. For instance, for a lattice size $96 \times 96$ in the $2d$ Ising model one has $9217$ possible energy values with $g(E)$ spanning over about $10^{4}$ orders of magnitude. The smallest polynomial coefficient is $g(-18432)=2$, the largest one is  $ g(0)=2.304\times 10^{2772}$! An example of the difficulties found by the zeros finder is seen in Fig.~\ref{Fig1}. To find those zeros we have used the package \emph{solve} of the Mathematica\textsuperscript{\textregistered} software. It is evident that the symmetry is not preserved. At this point, one may be tempted to consider only the \emph{most relevant} coefficients of the polynomial, probably the largest ones. However, these coefficients are generally related to very high energy states, unlikely to play any important role in the phase transition. Then, it should be interesting if we could ``filter'' the region where the dominant zero is located and still have the relevant information about the phase transition. This, in fact, is possible to be done in a systematic way as we will show in the following.

\section{Energy Probability Distribution Zeros}

    To introduce the method note that if we multiply Eq.~\ref{Equation1} by $1 = e^{-\beta_0 \textbf{E}} e^{\beta_0 \textbf{E}}$ it can be rewritten as
 \begin{equation}
    Z_{\beta_0} = \sum_{\textbf{E}} h_{\beta_0} \left( \textbf{E} \right) e^{-\textbf{E} \Delta \beta},
\end{equation}
\noindent
where $h_{\beta_0} \left( \textbf{E} \right)=g \left( \textbf{E} \right)e^{-\beta_0 E}$ and $\Delta \beta = \beta - \beta_0$. Defining the variable $x = e^{- \Delta \beta \varepsilon }$ we get the partition function
\begin{equation}
    Z_{\beta_0} = e^{-  \Delta \beta  \varepsilon_0} \sum_{n} h_{\beta_0}({ \emph{n}}) x^n  ,
    \label{EPD_polynomial}
\end{equation}
\noindent
where $h_{\beta_0}({\emph{n}})= h_{\beta_0} \left( \textbf{E}_n \right)$ is nothing but the unnormalized canonical energy probability distribution (EPD), hereafter referred to as the energy histogram at temperature $\beta_0$. There is a one to one correspondence between the Fisher zeros and the EPD zeros since they are related through a trivial conformal transformation.

    Constructing the histogram at the transition temperature, i.e., $\beta_0 = \beta_c$, the dominant zero will be at $x_c = 1$, i.e., $Z=0$ at the critical temperature ($\Delta \beta =0$) in the thermodynamic limit. For finite systems a small imaginary part of $x_c$ is expected.  Indeed, we may expect that the dominant zero is the one with the smallest imaginary part on the real positive region regardless $\beta_0$. Once we locate the dominant zero its distance to the point $(1,0)$ gives $\Delta \beta$ and an estimate for $\beta_c$.

    For temperatures close enough to $\beta_c$ only states with a non-vanishing probability to occur are pertinent to the phase transition. Thus, for $\beta_0 \approx \beta_c$ we can judiciously discard \emph{small} values of $h_{\beta_0}$. As a consequence, the zeros finder does not have to deal with high degree polynomials with coefficients spanning over many orders of magnitude. Moreover, one may speculate that the dominant zero act as an accumulation point such that even far from $\beta_c$ fair estimates can be obtained.

\begin{figure}
    \includegraphics[keepaspectratio,height=0.8\textwidth]{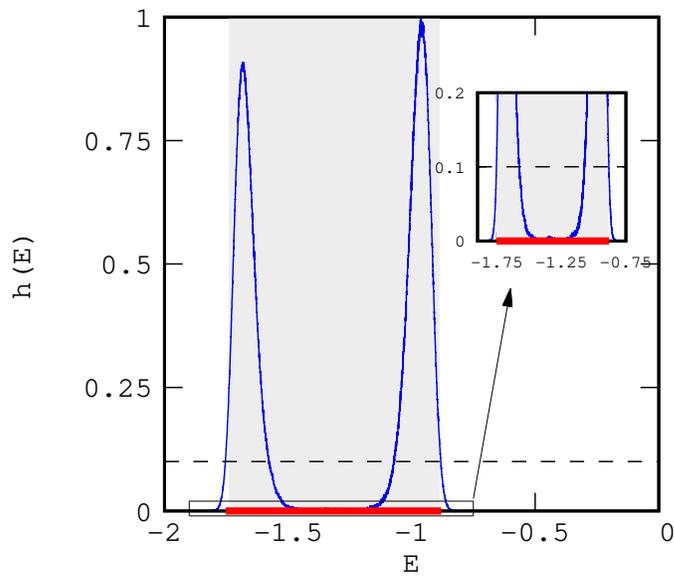}
        \\
      \vspace*{0.75cm}
    \caption{(Color online) Example of the criteria used to build the energy subspace to calculate the zeros distribution. The figure is for a typical first order transition. Coming from the \emph{outer} region we cut the low and high energy tails of the histogram using a cutoff $h_{cut}$ (in this figure $h_{cut}=0.1$) which defines $E_{min}$ and $E_{max}$. The intermediate region is preserved.}
    \label{Fig1.1}
\end{figure}

    With this in mind we can develop a clear criterion to filter the important region in the energy space were the most relevant zeros are located. The idea follows closely the well known \emph{Regula Falsi}  method for solving an equation in one unknown. The reasoning is as follows: We first build a rescaled histogram $ h_{\beta_0^0}$ ($\textrm{Max}(h_{\beta_0^{0}})=1$) at an initial (False) guess $\beta_{0}^{0}$ and reduce the considered energy range (polynomial degree) by discarding states with low enough values of $ h_{\beta_0^0}$, say $h_{\beta_0^0} < h_{cut}$ (more details below). Afterward, we construct the polynomial, Eq.~\ref{EPD_polynomial}, finding the corresponding zeros. By selecting the dominant zero, $x_c^0$, we can estimate the pseudo critical temperature, $\beta_c^0$. Regarding that $\beta_c(L)$ is the true pseudo-critical temperature for the system of size $L$, if the initial guess $\beta_0^0$ is far from $\beta_c(L)$ the estimative $\beta_c^0$ will not be satisfactory. Nevertheless, we can proceed iteratively making $\beta_0^1=\beta_c^0$, building a new histogram at this temperature and starting over. After a \emph{reasonable} number of iterations we may expect that $ \beta_{c}^{j} $ converges to the true $ \beta_c(L) $ and thus $x_c^j$ approaches the point $(1,0)$.  This corresponds to apply a sequence of transformations $\mathfrak{T}$ such that $\beta^{n+1} = \mathfrak{T} \beta^{n}$. The transition temperature corresponds to the fixed point $\beta_c = \mathfrak{T} \beta_c$. The property $x_c^j \to (1,0)$ can be used as a consistency check in this iterative process. An algorithm following those ideas is:
\begin{enumerate}
    \item{Build a single histogram $ h_{\beta_0^j}$ at $\beta_0^j$.}
    \item{Find the zeros of the polynomial with coefficients given by $ h_{\beta_0^j}$.}
    \item{Find the dominant zero, $ x_c^j$.}
	\subitem a) {If $ x_c^j$ is close enough to the point $(1,0)$, stop.}
	\subitem b) {Else, make $\beta_0^{j+1}=-\frac{\ln \left (\Re e\left \{x_c^j \right \} \right ) }{\varepsilon}+\beta_0^j$ and go back to 1.}
\end{enumerate}
\noindent
    In all our numerical results it seems that the choice of the starting temperature is irrelevant. To build the single histogram we follow the recipe given by Ferrenberg and Swendsen \cite{Ferrenberg1,Ferrenberg2}. To specify the energy subspace, the procedure is straightforward dealing with continuous transitions. In a first order transition the probability distribution is bi-modal. The energy distribution between the two peaks may reach low values, even lower than our cutoff criteria ($h_{cut}$). In this case the procedure to define the relevant energy interval is subtler. A general way for dealing with this is to cut the low and high energy tails of the probability density preserving the important intermediate region as shown in Fig. \ref{Fig1.1}. Of course, different choices for the cutoff, $h_{cut}$, will affect the final results accuracy. Small values for $h_{cut}$ are expected to increase accuracy while making the zeros finder task more difficult. The final choice is a compromise between the desired method accuracy and zeros finder accuracy.
\noindent
\section{Testing the 2D Ising Model}

\begin{table}
    \caption{Estimates of the pseudo critical temperature for the ferromagnetic 2d Ising Model. The last line is the exact value up to five figures.}
\begin{tabular}{ccccc}
  \hline
  L  & $\beta$       & $T$        & $|x|$       & $\Im m(x)$     \\
  \hline \hline
  4  & 0.6667        & 3.0000     & 0.6262             & 0.7652            \\
     & 0.4504        & 2.2202     & 1.1130             & 1.2180            \\
     & 0.4236        & 2.3607     & 0.9752             & 1.0939             \\
     & 0.4299        & 2.3261     & 1.0059             & 1.1218             \\
     & 0.4284        & 2.3343     & 0.9985             & 1.1150             \\
     & 0.4288        & 2.3321     & 1.00045            & 1.1168             \\
     & 0.4287        & 2.3326     & 0.9999             & 1.1159             \\
  \hline
  8  & 0.4287        & 2.3326     & 0.9924             & 0.4241             \\
     & 0.4306        & 2.3223     & 0.9999             & 0.4273             \\
  \hline
  16 & 0.4306        & 2.3223     & 0.9923             & 0.2088             \\
     & 0.4325        & 2.3121     & 0.9999             & 0.2104             \\
  \hline
  32 & 0.4325        & 2.3121     & 0.9872             & 0.1035             \\
     & 0.4357        & 2.2952     & 0.9998             & 0.1049             \\
     & 0.43575       & 2.2949     & 0.9999             & 0.1049             \\
  \hline
  64 & 0.43575       & 2.2949     & 0.9911             & 0.0520             \\
     & 0.4380        & 2.2831     & 1.00005            & 0.0525             \\
  \hline
  96 & 0.43575       & 2.2949     & 0.9967             & 0.0349             \\
     & 0.4388        & 2.2789     & 0.9999             & 0.0350             \\
  \hline \hline
  estimate  & 0.4404       & 2.2707     & -            & -                   \\
  \hline
 $\infty$   & 0.4407       & 2.2691     & -            & -                   \\
  \hline   \hline
\end{tabular}
\label{Table1}
\end{table}
\noindent
\begin{figure}
    \includegraphics[keepaspectratio,height=0.8\textwidth]{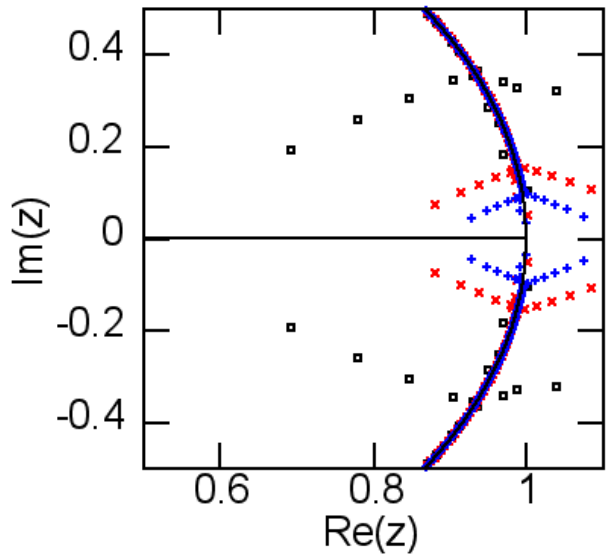}
            \\
      \vspace*{0.75cm}
    \caption{Plot section of the zeros distribution for the Ising model. Lattices sizes are $L=32,64,96$ ($\Box , \times$ and $+$ respectively.). The solid black line marks the perimeter of the unit circle and the real axis.}
\label{Fig2}
\end{figure}
\noindent
     The convergency and accuracy of the method can be tested by using the well known Ising model in a square lattice defined by the following Hamiltonian
\begin{equation}
    H=-J\sum_{<i,j>}\sigma_i\sigma_j,
\end{equation}
\noindent
    where $J$ is a ferromagnetic coupling constant, $<i,j>$ means that the summation runs over nearest neighbors on a regular lattice and $\sigma_i=\pm1$ is a classical spin variable at site $i$. The model has a continuous phase transition at $T_c=\frac{2}{\ln(1+\sqrt{2})}\frac{J}{k_B}\approx 2.269\frac{J}{k_B}$ and was exactly solved in 1944~\cite{Onsager1944}, being thus one of the preferred model for testing new methods on phase transitions. Hereafter we'll consider $J=1$ and $k_B=1$.

\begin{figure}
    \includegraphics[keepaspectratio,height=0.8\textwidth]{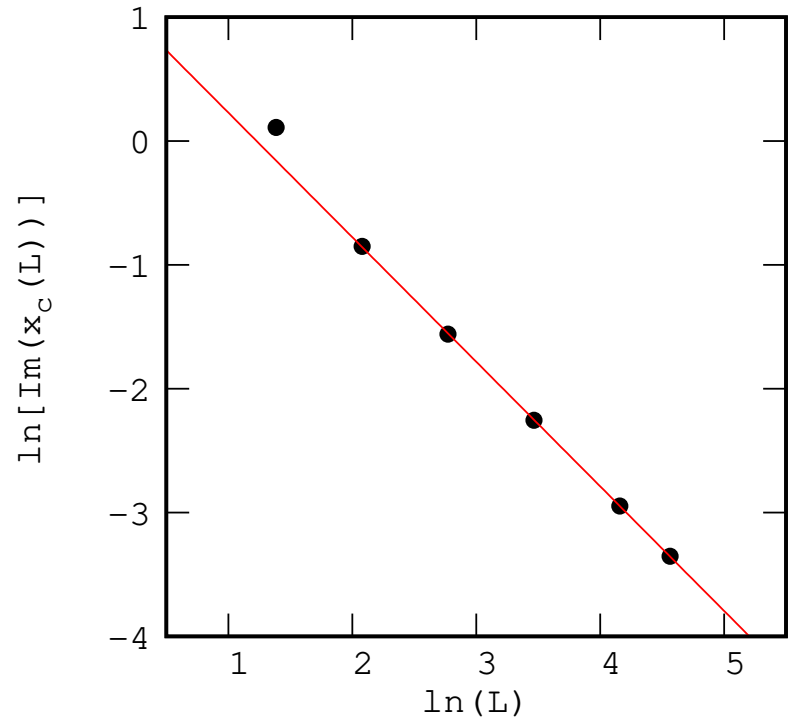}
            \\
      \vspace*{1.75cm}
    \caption{Log-log plot of $\Im m(x_c(L))$ versus $L$ for the Ising model. According to finite size scaling theory we expect the slope to be $1/\nu$. From our data we obtained $1/\nu=1.006(3)$ discarding the point corresponding to $L=4$.}
\label{fig1-ising}
\end{figure}

    In Tab.~\ref{Table1} we show the results of the application of our method to the Ising model in a square lattice for several lattice sizes ($4\leq L \leq 96$). The results were obtained by using the exact density of states calculated using the Mathematica\textsuperscript{\textregistered} software with the code developed in Ref.~\cite{exact-dos-ising}. The interval of energy used was $[E_{min}, E_{max}]$ chosen in such way that for the rescaled histogram $h_{\beta}({\emph{n}}) > h_{cut}=10^{-2}$ (see Fig.~\ref{Fig1.1}). The zeros of the polynomial were calculated using the Mathematica\textsuperscript{\textregistered} function \emph{Solve}. In the last two lines of Tab.~\ref{Table1} we show the critical temperature result obtained from finite size scaling and the exact value up to $5$ figures. The error is smaller than  0.5\%. The zeros distribution is shown in Fig.~\ref{Fig2}. Another important quantity that can readily be obtained is the critical exponent $\nu $. Finite size scaling theory predicts that $T_c(L)\sim T_c + aL^{-1/\nu}$~\cite{privman1990} and thus, we might expect that $x_c(L) \sim x_c+bL^{-1/\nu}$ \cite{Itzykson83}. Once $\Re e\{x_c(L)\}\approx 1, \quad \forall L$, the imaginary part of $x_c(L)$ should also scale as $L^{-1/\nu}$. In fig.~\ref{fig1-ising} is shown a log-log plot of $\Im m(x_c(L))$ versus $L$ for $L= 4, 8, 16, 32, 64, 96$. Considering all points we obtain $1/\nu=1.07(3)$. Discarding the smaller lattice size gives $1/\nu=1.006(3)$, very close to the exact value, $1$. Of course, better results for $T_c$ and $\nu$ can be obtained by diminishing the energy threshold ($h_{cut}$) and using larger lattice sizes.
\noindent
\section{2D Potts Model}
    The $q$ states Potts Model~\cite{Wu82} is defined by the Hamiltonian,
\begin{equation}
    H=-J\sum_{<i,j>}\delta_{\sigma_i,\sigma_j},
\end{equation}
\noindent
    where $J$ is a ferromagnetic coupling constant, $<i,j>$ stands for nearest neighbors, $\delta_{\alpha,\beta}$ is the Kronecker delta ($\delta_{\alpha\beta}=1$ if $\alpha = \beta$, 0 otherwise) and $\sigma_i = 1,2,\cdots,q$ is a spin like variable. The model can undergo a continuous (second order) transition for $q\le4$ and a discontinuous (first order) phase transition for $q>4$ at $T_c=\frac{1}{\ln(1+\sqrt{q})}\frac{J}{k_B}$.  We have simulated the model in a square lattice of linear size $L$. One Monte Carlo Step (MCS) constitutes the attempt to assign a new value to $N=L \times L$ spins. The only quantity of interest in our simulations is the energy histogram, which is used to obtain the zeros. After discarding $N_{term}$ MCS for thermalization the energy histogram was built using $N_{MCS}$ MCS. The threshold, $ h_{cut}$, was $10^{-3}$. In what follows we detail the calculations for each $q$ value studied.
\subsection{3 states}
\begin{figure}
    \includegraphics[keepaspectratio,height=0.8\textwidth]{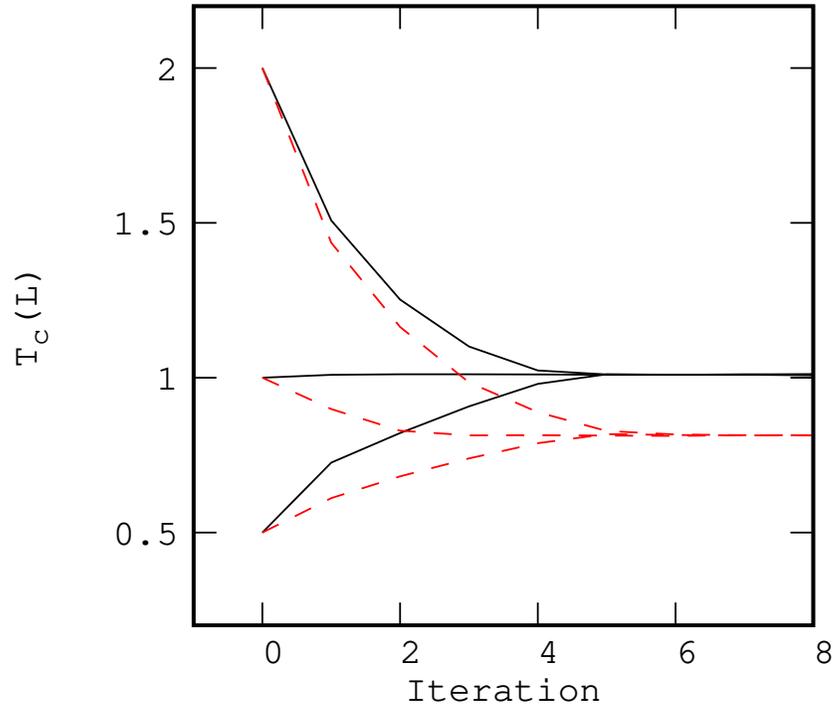}
            \\
      \vspace*{1.75cm}
    \caption{Estimated critical temperature at each iteration for the 3 (full black lines) and 6 (dashed red lines) states Potts models for a $L=20$ square lattice showing the fast convergence to the pseudo-critical temperature.}
\label{Fig3}
\end{figure}

\noindent
    Initial scans for the critical temperature were done for $L=20$ using $N_{term}=50,000$ and $N_{MCS}=500,000$. To show the final estimate independence of the initial guess we used three different  starting temperatures $T_0^1=2.0$, $1.0$ and $0.5$. After approximately $6$ iterations the temperature converged to $T_c(20)\approx 1.009$, as shown in fig.~\ref{Fig3}. This value is then used as the initial temperature for another set of simulations. Table~\ref{table_q3} shows detailed information about the simulation temperatures ($T_{sim}$) and parameters for each lattice size. Errors were estimated by considering statistical fluctuations among $5$ independent Monte Carlo simulations for each lattice size. In order to get precise results we iterate our algorithm for each lattice size and sample until $|\Re e\{ {x_c(L)} \}-1| < 10^{-4}$.
\noindent
\begin{table}
    \caption{Monte Carlo parameters ($N_{term}$, $N_{MCS}$ and $T_{sim}$) and results ($\Im m(x_{c}(L))$ and $T_c(L)$) for the 3 states Potts model.}
    \begin{tabular}{cccccc}
  \hline
 L  & $N_{term}$       & $N_{MCS}$        & $T_{sim}$     & $\Im m(x_c(L))$ & $T_c(L)$     \\
  \hline \hline
    20 & $10^5$ & $5\times 10^6$ & 1.0090 & $5.151(9)\times 10^{-2}$ & 1.00978(3) \\ \hline
    40 & $10^5$ & $5\times 10^6$ & 1.0000 & $2.26(1) \times 10^{-2}$ & 1.0016(1) \\ \hline
    60 & $3\times10^5$ & $5\times 10^6$ & 0.9995 & $1.42(2)\times 10^{-2}$ & 0.9992(1) \\ \hline
    80 & $3\times10^5$ & $5\times 10^6$ & 0.9970 & $1.00(1)\times 10^{-2}$ & 0.9979(2)  \\ \hline
    120 & $3\times10^5$ & $5\times 10^6$ & 0.9965 & $6.157(5)\times 10^{-3}$ & 0.99663(5) \\ \hline
    160 & $3\times10^5$ & $5\times 10^6$ & 0.9960 & $4.289(4)\times 10^{-3}$ & 0.9960(1) \\ \hline \hline
\end{tabular}
\label{table_q3}
\end{table}
\noindent

\begin{figure}
    \includegraphics[keepaspectratio,height=0.8\textwidth]{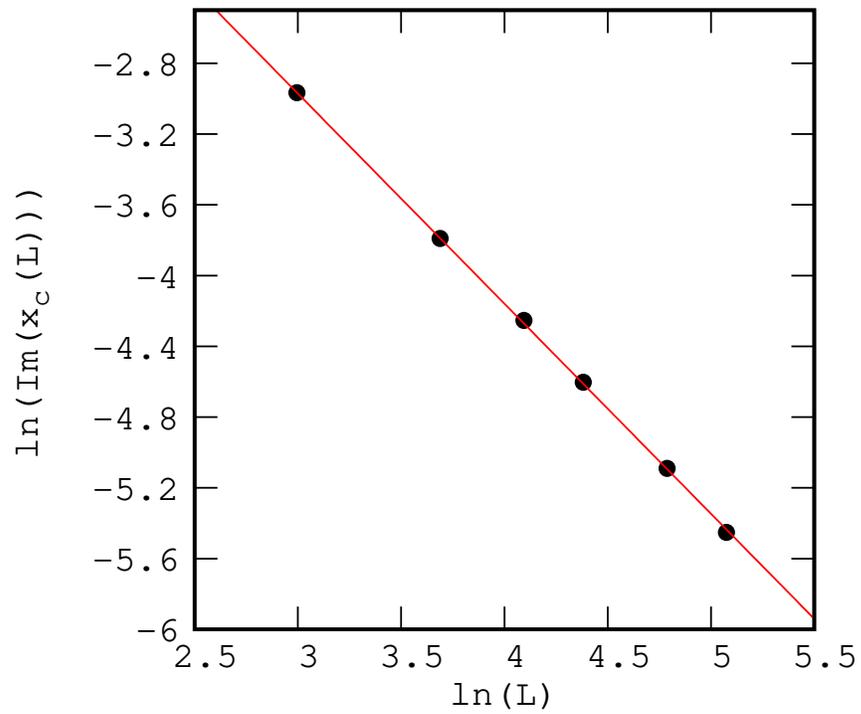}
            \\
      \vspace*{1.75cm}
    \caption{Log-log plot of $\Im m(x_{c}(L))$ versus $L$ for the 3 states Potts model. According to finite size scaling theory we expect the slope to be $1/\nu$. From our data we obtained $1/\nu=1.19(1)$. When not shown error bars are smaller than the symbol size.}
\label{fig1-q3}
\end{figure}

\begin{figure}
    \includegraphics[keepaspectratio,height=0.8\textwidth]{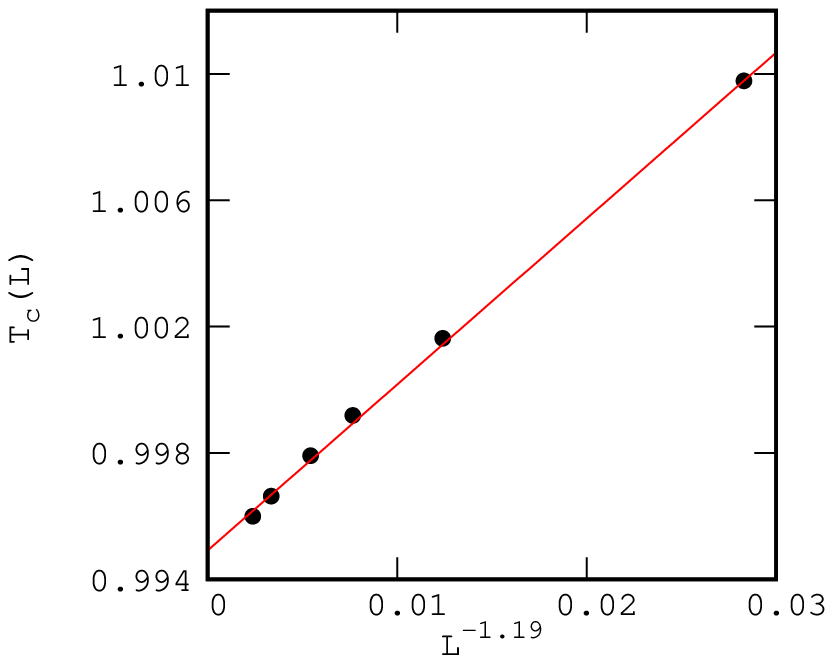}
            \\
      \vspace*{1.75cm}
    \caption{$T_c(L)$ versus $L^{-1/\nu}$ for the 3 states Potts model. Extrapolation to thermodynamical limit ($L\to \infty$) gives $T_c(\infty)=0.99491(7)$. When not shown error bars are smaller than the symbol size.}
\label{fig2-q3}
\end{figure}

    In figs.~\ref{fig1-q3} and~\ref{fig2-q3} we show the finite size scaling analysis for the determination of the critical exponent and temperature respectively. In the first, a log-log plot of the imaginary part of the leading zero as a function of the lattice size gives an exponent $1/\nu=1.19(1)$, in excellent agreement  with the exact value, $6/5=1.2$~\cite{Wu82}. Using this value, we plotted $T_c(L)$ versus $L^{-1/\nu}$ to obtain $T_c(\infty)=0.99491(7)$, again, in excellent agreement  with the exact value, $0.99497$.
\noindent
\noindent
\subsection{6 states}
\begin{figure}
    \includegraphics[keepaspectratio,height=0.8\textwidth]{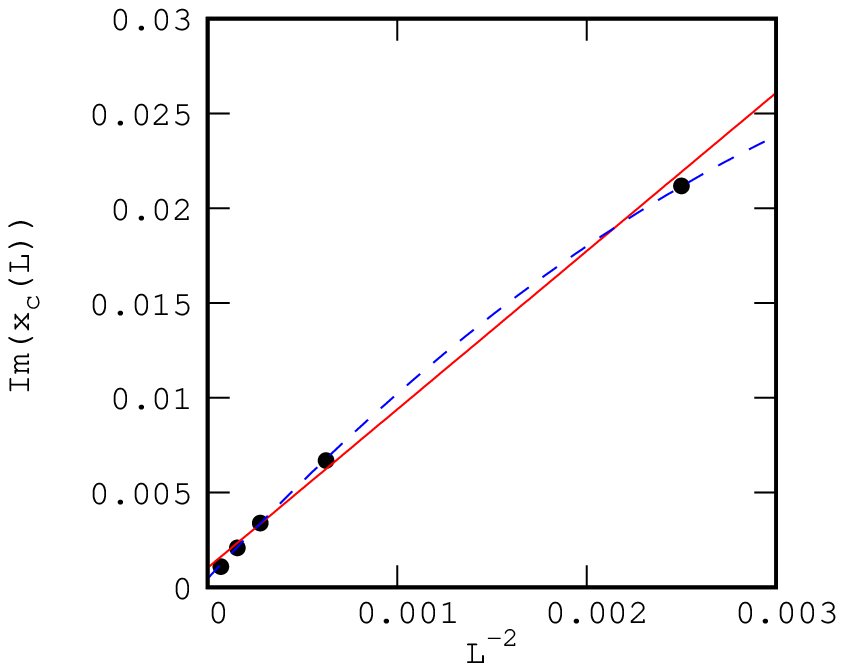}
            \\
      \vspace*{1.75cm}
    \caption{$\Im m(x_c(L))$ versus $L^{-2}$ for the 6 states Potts model. The solid red line is a linear adjust wile dashed blue line is a quadratic fit. As can be seen with the quadratic fit, i.e., when corrections to scaling are considered, the data are very well adjusted. When not shown error bars are smaller than the symbol size.}
\label{fig1-q6}
\end{figure}

\begin{figure}
    \includegraphics[keepaspectratio,height=0.8\textwidth]{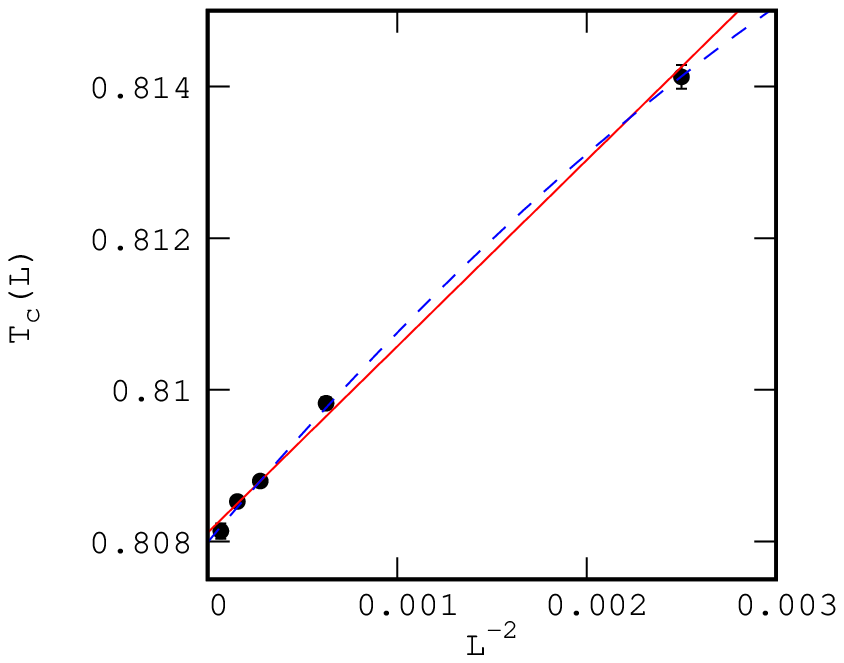}
            \\
      \vspace*{1.75cm}
    \caption{$T_c(L)$ versus $L^{-2}$ for the 6 states Potts model. Solid line is a linear adjust while dashed line is a quadratic fit. As can be seen with the quadratic fit, i.e., when corrections to scaling are considered, the data are very well adjusted. From the adjust we get $0.80797(7)$. When not shown error bars are smaller than the symbol size.}
\label{fig2-q6}
\end{figure}

    For the $6$ states model we used the same procedure as before. To refine the final results we have used the multiple histogram reweighting~\cite{Ferrenberg2} technique, which allowed a better sampling near the weak first order transition. Table~\ref{table-q6} shows in detail our results.
\begin{table}
    \caption{Monte Carlo parameters ($N_{term}$, $N_{MCS}$ and $T_{sim}$) and results ($\Im m(x_{c}(L))$ and $T_c(L)$) for the $6$ states Potts model.}
\begin{tabular}{cccccc}
  \hline
  L  & $N_{term}$       & $N_{MCS}$        & $T_{sim}$     & $\Im m(x_c(L))$ & $T_c(L)$     \\
  \hline \hline
20 & $10^5$ & $5\times 10^6$ & 0.8100&  0.2118(7) &  0.814(2)\\ \hline
40 & $10^5$ & $5\times 10^6$ & 0.8100 & 0.669(3) &  0.8098(8) \\ \hline
60 & $3\times10^5$ & $5\times 10^6$ & 0.810 &  &  \\
     & $3\times10^5$  & $5\times 10^6$ & 0.809 &  &  \\
          &  $3\times10^5$  & $5\times 10^6$ & 0.808 & 0.3387(8) &  0.80880(5) \\ \hline
80 & $3\times10^5$ & $5\times 10^6$ & 0.810 &  &   \\
     & $3\times10^5$  & $5\times 10^6$  & 0.809 &  &  \\
          &  $3\times10^5$   &  $5\times 10^6$  & 0.808 & 0.208(3) & 0.8085(5) \\ \hline
120 & $3\times10^5$ & $5\times 10^6$ & 0.810 &  &  \\
     &  $3\times10^5$ &  $5\times 10^6$ & 0.809 &  &  \\
     & $3\times10^5$ &  $5\times 10^6$ & 0.808 &  &  \\
     & $3\times10^5$  &  $5\times 10^6$ & 0.807 & 0.109(3) &  0.808(1) \\ \hline \hline
\end{tabular}
\label{table-q6}
\end{table}
\noindent
    The finite size scaling in this case has shown to be more subtle. For this weak first order transition the scaling functions need to be considered in more details~\cite{privman1990}. In Figs. ~\ref{fig1-q6} and~\ref{fig2-q6} are shown both the imaginary part of $x_c(L)$ and the pseudo-critical temperature as a function of $L^{-d}$. A linear adjust does not work in this case, anticipating that a correction is missing in the scaling function. Following Lee and Kostelitz \cite{Lee-Kosterlitz}  we introduce a $L^{-2d}$ scaling correction which fits quite well our results as seen in figs.~\ref{fig1-q6} and~\ref{fig2-q6}. The critical temperature encountered ($0.80797(7)$) agrees up to $5$ figures with the exact value ($0.80761$).
\noindent
\noindent
\section{Anisotropic Heisenberg or XY Model}

\begin{figure}
    \includegraphics[keepaspectratio,height=0.8\textwidth]{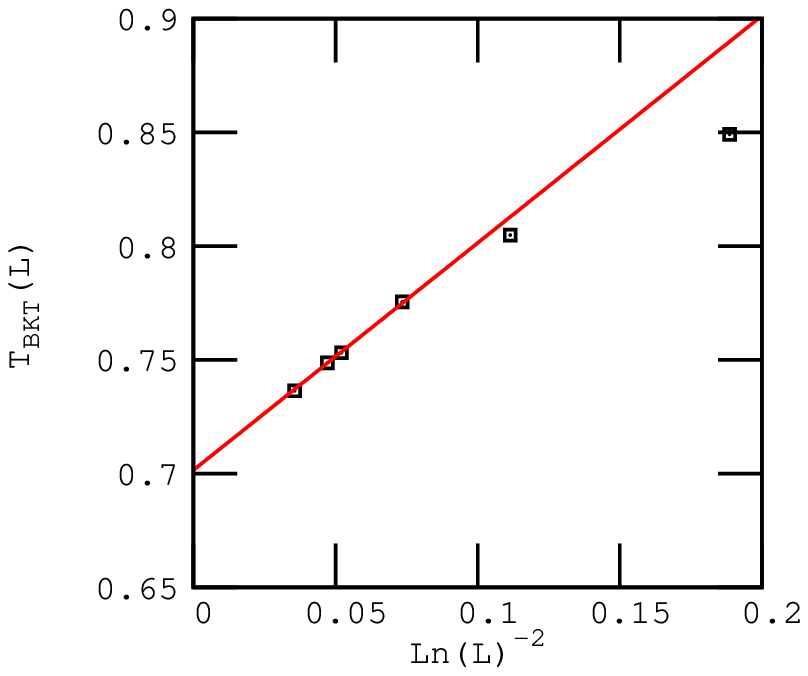}
            \\
      \vspace*{1.75cm}
    \caption{$T_{BKT}(L)$ versus $1/\ln(L)^{2}$ for the XY model. The solid line is a linear adjust from which we get $T_c=0.7003(3)$. Small lattice sizes ($L=10$ and 20) were not considered in the adjust. When not shown error bars are smaller than the symbol size. }
\label{fig1-xy}
\end{figure}

    Another test to our algorithm is the  $2d$ anisotropic Heisenberg model, also known as $XY$ model, which has a $BKT$ transition~\cite{B,KT}. In this model the spin variable at site $i$ is a classical 3D vector, $\vec{S}_i=(S^x_i,S^y_i, S^z_i)$, which interacts with first neighbors as
\begin{equation}
    H=-J\sum_{<i,j>} \left( S^{x}_{i} S^{x}_{j} + S^{y}_{i} S^{y}_{j} \right).
\end{equation}
\noindent
      For this model we have used the Replica-Exchange-Wang-Landau algorithm\cite{Wang-Landau,Vogel} to estimate the histograms. Details of its implementation can be found in reference Ref.~\cite{Julio16}. This model has a continuum energy spectrum. To overcome this we have used a discretized energy spectrum $E_n = n \varepsilon+\varepsilon_0$ with $\varepsilon=1J$~\cite{Polymer13,Julio16}. The last column in Tab.~\ref{Table3} shows the final results averaged over $5$ independent simulations. The last line is the result of the finite size scaling analysis. Following Kenna~\cite{Kenna_2} the $BKT$ transition temperature scales as $T_{BKT}(L) \sim [\ln(L)]^{-2}$. Our results are shown in Fig.~\ref{fig1-xy}. We have obtained $T_{BKT}\approx0.7003(3)$ which agree quite well with the estimate $T_{BKT}\approx0.700$ obtained in several other works~\cite{Evertz,PLA_Costa,Julio16})
\begin{table}
    \caption{Estimate of the $BKT$ transition temperature for the XY Model.}
\begin{tabular}{cc}
  \hline
  L  & $T_{BKT}(L)$   \\
  \hline \hline
  10 & 0.8491(1)      \\
  \hline
  20  & 0.8049(5)     \\
  \hline
  40  & 0.7755(6)     \\
  \hline
  80  & 0.7536(4)     \\
  \hline
  100 & 0.7587(2)     \\
  \hline
  200 & 0.7364(3)     \\
  \hline  \hline
  $<T_{BKT}>$ & 0.7003(3)  \\
    \hline  \hline
\end{tabular}
\label{Table3}
\end{table}
\noindent
\noindent
\section{13 monomer homopolymer Model}
    As a final test to our algorithm we considered a $13$ monomer flexible homopolymer, which is an intrinsically finite system with two pseudo-transitions, a first order transition at $T_{1st}\approx 0.33520$ followed by a continuous  transition at $T_{2nd} \approx 1.1292$~\cite{Polymer13}. In this model, we considered that unbounded monomers interacts via a truncated and shifted Lennard-Jones potential,
    \begin{equation}
V_{LJ}^{mod}(r_{ij})=V_{LJ}[\textrm{min}(r_{ij},r_c)]-V_{LJ}(r_c),
    \end{equation}
    where $r_{ij}$ is the distance between monomers $i$ and $j$, $r_c$ is the cutoff distance and
    \begin{equation}
    V_{LJ}(r)=4\varepsilon_{LJ}\left [  \left ( \frac{\sigma}{r} \right )^{12} - \left ( \frac{\sigma}{r} \right )^{6} \right ],
    \end{equation}
    \noindent
    is the standard Lennard-Jones potential. We choose $\varepsilon_{LJ}=1$, $\sigma = 2^{-1/6} r_0$ and $r_c=2.5\sigma$. The elastic bonds between adjacent monomers is modeled by the finitely extensible nonlinear elastic (FENE) potential,
    \begin{equation}
    V_{FENE}(r_{i,i+1})=-\frac{K}{2}R^2 \ln \left[ 1- \left( \frac{r_{i,i+1}-r_0}{R} \right )^2 \right ].
    \end{equation}
    \noindent
    This potential possesses a minimum at $r_0$ and diverges for $r \to r_0 \pm R$. $K$ is a spring constant. We follow Ref.~\cite{Polymer13} to set the parameters to $R=0.3$, $r_0 =0.7$, and $K=40$. The histograms were obtained from the density of states by using the conventional Wang-Landau algorithm~\cite{Wang-Landau}. The flatness criteria was set to $0.7$ and the final actualization factor was $\ln f = 10^{-9}$. A discretized set of energies with $\varepsilon=1$ was used. Our results are in complete accordance with those obtained in Ref.~\cite{Polymer13} and are depicted in Tab.~\ref{Table2}. It is noteworthy that if we start the iteration at $T_{0} > T_{2nd}$ or $T_{0} < T_{1st}$ they always converge to $T_{2nd}$ or $T_{1st}$ respectively. If we use an intermediate value it can converge to one or another. The borders of the basin of convergence were not studied in this work. The first order character of the transition at $T_{1st}$ is evident in the double peak structure of the histogram (not shown here).
\begin{table}
    \caption{Determination of the transition temperatures for the $13$ monomers elastic and flexible polymer. Two transitions are found. A low temperature first order and a second at higher temperature. The transition temperatures are to be compared with those in Ref.~\cite{Polymer13}. }
\begin{tabular}{ccccc}
  \hline
  L  & $\beta$       & $T$        & $|x|$       & $\Im m(x)$         \\
  \hline \hline
  13 & 10.0000       & 0.10000     & 1.02230            & 0.02650            \\
     & 6.60690       & 0.15140     & 1.01414            & 0.01660            \\
     & 4.44675       & 0.22488     & 1.00855            & 0.00806            \\
     & 3.13696       & 0.31878     & 1.00135            & 0.00502            \\
     & 2.92940       & 0.34142     & 0.99965            & 0.00502            \\
     & 2.98685       & 0.33520     & 1.00002            & 0.00502            \\
  \hline
    13 & 0.8333        & 1.2000      & 0.99988            & 0.00425            \\
     & 0.8487        & 1.1783      & 0.99991            & 0.00423            \\
     & 0.8625        & 1.1594      & 0.99994            & 0.00421            \\
     & 0.8717        & 1.1472      & 0.99996            & 0.00420            \\
     & 0.8779        & 1.1391      & 0.99997            & 0.00420            \\
     & 0.8825        & 1.1331      & 0.99998            & 0.00420            \\
     & 0.8856        & 1.1292      & 0.99999            & 0.00419            \\
  \hline \hline
\end{tabular}
\label{Table2}
\end{table}
\noindent
\section{Closing Remarks}

    In summary, we introduced a novel method to study phase transitions based on the zeros of the energy probability distribution. The method has shown to be fully applicable to any model allowing accurate determination of the critical temperatures and the exponent $\nu$ as demonstrated by our results in the Ising, Potts, Heisenberg and polymer models. The only criteria we have used was the fact that at the transition temperature the partition function zero goes to the real positive axis in the thermodynamic limit. No \emph{a priori}  knowledge of an order parameter is necessary. The method is iterative, so that, the transition temperature can be approached at will. Besides that, the knowledge of the complete density of states is not required since we work in a restricted range of energy. The advantage is that the polynomial representing the partition function has fewer roots with its coefficients ranging in a civilized, narrow region. Other advantage of the method is that ambiguities in the determination of pseudo-critical temperatures existing in canonical analysis in intrinsically finite systems is removed as evidenced by the polymer model studied here.
\section*{Acknowledgments}
    This work was partially supported by CNPq and Fapemig, Brazilian Agencies. JCSR thanks CNPq for the support under grant PDJ No. 150503/2014-8 and Grant CNPq 402091/2012-4.

\section*{References}

\bibliography{references}

\begin{thebibliography}{10}
\expandafter\ifx\csname url\endcsname\relax
  \def\url#1{\texttt{#1}}\fi
\expandafter\ifx\csname urlprefix\endcsname\relax\def\urlprefix{URL }\fi
\expandafter\ifx\csname href\endcsname\relax
  \def\href#1#2{#2} \def\path#1{#1}\fi

\bibitem{Kadanoff2009}
L.~P. Kadanoff, \href{http://dx.doi.org/10.1007/s10955-009-9814-1}{More is the
  same; phase transitions and mean field theories}, Journal of Statistical
  Physics 137~(5) (2009) 777--797.
\newblock \href {http://dx.doi.org/10.1007/s10955-009-9814-1}
  {\path{doi:10.1007/s10955-009-9814-1}}.
\newline\urlprefix\url{http://dx.doi.org/10.1007/s10955-009-9814-1}

\bibitem{Fisher1998}
M.~E. Fisher,
  \href{http://link.aps.org/doi/10.1103/RevModPhys.70.653}{Renormalization
  group theory: Its basis and formulation in statistical physics}, Rev. Mod.
  Phys. 70 (1998) 653--681.
\newblock \href {http://dx.doi.org/10.1103/RevModPhys.70.653}
  {\path{doi:10.1103/RevModPhys.70.653}}.
\newline\urlprefix\url{http://link.aps.org/doi/10.1103/RevModPhys.70.653}

\bibitem{Yang-Lee1952}
C.~N. Yang, T.~D. Lee,
  \href{http://link.aps.org/doi/10.1103/PhysRev.87.404}{Statistical theory of
  equations of state and phase transitions. i. theory of condensation}, Phys.
  Rev. 87 (1952) 404--409.
\newblock \href {http://dx.doi.org/10.1103/PhysRev.87.404}
  {\path{doi:10.1103/PhysRev.87.404}}.
\newline\urlprefix\url{http://link.aps.org/doi/10.1103/PhysRev.87.404}

\bibitem{Fisher1964}
M.~E. Fisher, in: W.~Brittin (Ed.), Lectures in Theoretical Physics: Volume VII
  C - Statistical Physics, Weak Interactions, Field Theory : Lectures Delivered
  at the Summer Institute for Theoretical Physics, University of Colorado,
  Boulder, 1964, no. v. 7, University of Colorado Press, Boulder, 1965.

\bibitem{Swendsen_Wang}
R.~H. Swendsen, J.-S. Wang,
  \href{http://link.aps.org/doi/10.1103/PhysRevLett.58.86}{Nonuniversal
  critical dynamics in monte carlo simulations}, Phys. Rev. Lett. 58 (1987)
  86--88.
\newblock \href {http://dx.doi.org/10.1103/PhysRevLett.58.86}
  {\path{doi:10.1103/PhysRevLett.58.86}}.
\newline\urlprefix\url{http://link.aps.org/doi/10.1103/PhysRevLett.58.86}

\bibitem{Wolff}
U.~Wolff, \href{http://link.aps.org/doi/10.1103/PhysRevLett.62.361}{Collective
  monte carlo updating for spin systems}, Phys. Rev. Lett. 62 (1989) 361--364.
\newblock \href {http://dx.doi.org/10.1103/PhysRevLett.62.361}
  {\path{doi:10.1103/PhysRevLett.62.361}}.
\newline\urlprefix\url{http://link.aps.org/doi/10.1103/PhysRevLett.62.361}

\bibitem{Muca1}
B.~A. Berg, T.~Neuhaus,
  \href{http://www.sciencedirect.com/science/article/pii/037026939191256U}{Multicanonical
  algorithms for first order phase transitions}, Physics Letters B 267~(2)
  (1991) 249 -- 253.
\newblock \href
  {http://dx.doi.org/http://dx.doi.org/10.1016/0370-2693(91)91256-U}
  {\path{doi:http://dx.doi.org/10.1016/0370-2693(91)91256-U}}.
\newline\urlprefix\url{http://www.sciencedirect.com/science/article/pii/037026939191256U}

\bibitem{Muca2}
B.~A. Berg, T.~Neuhaus,
  \href{http://link.aps.org/doi/10.1103/PhysRevLett.68.9}{Multicanonical
  ensemble: A new approach to simulate first-order phase transitions}, Phys.
  Rev. Lett. 68 (1992) 9--12.
\newblock \href {http://dx.doi.org/10.1103/PhysRevLett.68.9}
  {\path{doi:10.1103/PhysRevLett.68.9}}.
\newline\urlprefix\url{http://link.aps.org/doi/10.1103/PhysRevLett.68.9}

\bibitem{Muca3}
J.~Lee, \href{http://link.aps.org/doi/10.1103/PhysRevLett.71.211}{New monte
  carlo algorithm: Entropic sampling}, Phys. Rev. Lett. 71 (1993) 211--214.
\newblock \href {http://dx.doi.org/10.1103/PhysRevLett.71.211}
  {\path{doi:10.1103/PhysRevLett.71.211}}.
\newline\urlprefix\url{http://link.aps.org/doi/10.1103/PhysRevLett.71.211}

\bibitem{Muca4}
B.~A. Berg, \href{http://dx.doi.org/10.1007/BF02189233}{Multicanonical
  recursions}, Journal of Statistical Physics 82~(1) (1996) 323--342.
\newblock \href {http://dx.doi.org/10.1007/BF02189233}
  {\path{doi:10.1007/BF02189233}}.
\newline\urlprefix\url{http://dx.doi.org/10.1007/BF02189233}

\bibitem{Muca5}
B.~A. Berg,
  \href{http://www.sciencedirect.com/science/article/pii/S0920563297009626}{Algorithmic
  aspects of multicanonical simulations}, Nuclear Physics B - Proceedings
  Supplements 63~(1–3) (1998) 982 -- 984, proceedings of the \{XVth\}
  International Symposium on Lattice Field Theory.
\newblock \href
  {http://dx.doi.org/http://dx.doi.org/10.1016/S0920-5632(97)00962-6}
  {\path{doi:http://dx.doi.org/10.1016/S0920-5632(97)00962-6}}.
\newline\urlprefix\url{http://www.sciencedirect.com/science/article/pii/S0920563297009626}

\bibitem{Murilo1}
{de Oliveira, P. M.C.}, {Penna, T. J.P.}, {Herrmann, H. J.},
  \href{http://dx.doi.org/10.1007/s100510050172}{Broad histogram monte carlo},
  Eur. Phys. J. B 1~(2) (1998) 205--208.
\newblock \href {http://dx.doi.org/10.1007/s100510050172}
  {\path{doi:10.1007/s100510050172}}.
\newline\urlprefix\url{http://dx.doi.org/10.1007/s100510050172}

\bibitem{Murilo2}
{de Oliveira, P. M.C.}, \href{http://dx.doi.org/10.1007/s100510050532}{Broad
  histogram relation is exact}, Eur. Phys. J. B 6~(1) (1998) 111--115.
\newblock \href {http://dx.doi.org/10.1007/s100510050532}
  {\path{doi:10.1007/s100510050532}}.
\newline\urlprefix\url{http://dx.doi.org/10.1007/s100510050532}

\bibitem{Wang-Landau}
F.~Wang, D.~P. Landau,
  \href{http://link.aps.org/doi/10.1103/PhysRevLett.86.2050}{Efficient,
  multiple-range random walk algorithm to calculate the density of states},
  Phys. Rev. Lett. 86 (2001) 2050--2053.
\newblock \href {http://dx.doi.org/10.1103/PhysRevLett.86.2050}
  {\path{doi:10.1103/PhysRevLett.86.2050}}.
\newline\urlprefix\url{http://link.aps.org/doi/10.1103/PhysRevLett.86.2050}

\bibitem{Vogel}
T.~Vogel, Y.~W. Li, T.~W\"ust, D.~P. Landau,
  \href{http://link.aps.org/doi/10.1103/PhysRevLett.110.210603}{Generic,
  hierarchical framework for massively parallel wang-landau sampling}, Phys.
  Rev. Lett. 110 (2013) 210603.
\newblock \href {http://dx.doi.org/10.1103/PhysRevLett.110.210603}
  {\path{doi:10.1103/PhysRevLett.110.210603}}.
\newline\urlprefix\url{http://link.aps.org/doi/10.1103/PhysRevLett.110.210603}

\bibitem{Onsager1944}
L.~Onsager, \href{http://link.aps.org/doi/10.1103/PhysRev.65.117}{Crystal
  statistics. i. a two-dimensional model with an order-disorder transition},
  Phys. Rev. 65 (1944) 117--149.
\newblock \href {http://dx.doi.org/10.1103/PhysRev.65.117}
  {\path{doi:10.1103/PhysRev.65.117}}.
\newline\urlprefix\url{http://link.aps.org/doi/10.1103/PhysRev.65.117}

\bibitem{Wu82}
F.~Y. Wu, \href{http://link.aps.org/doi/10.1103/RevModPhys.54.235}{The potts
  model}, Rev. Mod. Phys. 54 (1982) 235--268.
\newblock \href {http://dx.doi.org/10.1103/RevModPhys.54.235}
  {\path{doi:10.1103/RevModPhys.54.235}}.
\newline\urlprefix\url{http://link.aps.org/doi/10.1103/RevModPhys.54.235}

\bibitem{B}
V.~L. Berezinskii, Journal of Experimental and Theoretical Physics - JETP 32
  (1971) 493.

\bibitem{KT}
J.~M. Kosterlitz, D.~J. Thouless,
  \href{http://stacks.iop.org/0022-3719/6/i=7/a=010}{Ordering, metastability
  and phase transitions in two-dimensional systems}, Journal of Physics C:
  Solid State Physics 6~(7) (1973) 1181.
\newline\urlprefix\url{http://stacks.iop.org/0022-3719/6/i=7/a=010}

\bibitem{Evertz}
H.~G. Evertz, D.~P. Landau,
  \href{http://link.aps.org/doi/10.1103/PhysRevB.54.12302}{Critical dynamics in
  the two-dimensional classical \textit{XY} model: A spin-dynamics study},
  Phys. Rev. B 54 (1996) 12302--12317.
\newblock \href {http://dx.doi.org/10.1103/PhysRevB.54.12302}
  {\path{doi:10.1103/PhysRevB.54.12302}}.
\newline\urlprefix\url{http://link.aps.org/doi/10.1103/PhysRevB.54.12302}

\bibitem{PLA_Costa}
B.~V. Costa, P.~Z. Coura, S.~A. Leonel,
  \href{http://www.sciencedirect.com/science/article/pii/S0375960113003150}{Berezinskii-kosterlitz-thouless
  transition close to the percolation threshold}, Physics Letters A 377~(18)
  (2013) 1239 -- 1241.
\newblock \href
  {http://dx.doi.org/http://dx.doi.org/10.1016/j.physleta.2013.03.030}
  {\path{doi:http://dx.doi.org/10.1016/j.physleta.2013.03.030}}.
\newline\urlprefix\url{http://www.sciencedirect.com/science/article/pii/S0375960113003150}

\bibitem{Schnabel2009}
S.~Schnabel, T.~Vogel, M.~Bachmann, W.~Janke,
  \href{http://www.sciencedirect.com/science/article/pii/S0009261409006186}{Surface
  effects in the crystallization process of elastic flexible polymers},
  Chemical Physics Letters 476~(4–6) (2009) 201 -- 204.
\newblock \href
  {http://dx.doi.org/http://dx.doi.org/10.1016/j.cplett.2009.05.052}
  {\path{doi:http://dx.doi.org/10.1016/j.cplett.2009.05.052}}.
\newline\urlprefix\url{http://www.sciencedirect.com/science/article/pii/S0009261409006186}

\bibitem{Schnabel2011}
S.~Schnabel, D.~T. Seaton, D.~P. Landau, M.~Bachmann,
  \href{http://link.aps.org/doi/10.1103/PhysRevE.84.011127}{Microcanonical
  entropy inflection points: Key to systematic understanding of transitions in
  finite systems}, Phys. Rev. E 84 (2011) 011127.
\newblock \href {http://dx.doi.org/10.1103/PhysRevE.84.011127}
  {\path{doi:10.1103/PhysRevE.84.011127}}.
\newline\urlprefix\url{http://link.aps.org/doi/10.1103/PhysRevE.84.011127}

\bibitem{Taylor2013}
M.~P. Taylor, P.~P. Aung, W.~Paul,
  \href{http://link.aps.org/doi/10.1103/PhysRevE.88.012604}{Partition function
  zeros and phase transitions for a square-well polymer chain}, Phys. Rev. E 88
  (2013) 012604.
\newblock \href {http://dx.doi.org/10.1103/PhysRevE.88.012604}
  {\path{doi:10.1103/PhysRevE.88.012604}}.
\newline\urlprefix\url{http://link.aps.org/doi/10.1103/PhysRevE.88.012604}

\bibitem{Julio16}
J.~C.~S. Rocha, L.~A.~S. M\'{o}l, B.~V. Costa,
  \href{http://www.sciencedirect.com/science/article/pii/S0010465516302466}{Using
  zeros of the canonical partition function map to detect signatures of a
  berezinskii-€"kosterlitz-€"thouless transition}, Computer Physics
  Communications 209 (2016) 88 -- 91.
\newblock \href {http://dx.doi.org/http://dx.doi.org/10.1016/j.cpc.2016.08.016}
  {\path{doi:http://dx.doi.org/10.1016/j.cpc.2016.08.016}}.
\newline\urlprefix\url{http://www.sciencedirect.com/science/article/pii/S0010465516302466}

\bibitem{exact-dos-ising}
P.~D. Beale, \href{http://link.aps.org/doi/10.1103/PhysRevLett.76.78}{Exact
  distribution of energies in the two-dimensional ising model}, Phys. Rev.
  Lett. 76 (1996) 78--81.
\newblock \href {http://dx.doi.org/10.1103/PhysRevLett.76.78}
  {\path{doi:10.1103/PhysRevLett.76.78}}.
\newline\urlprefix\url{http://link.aps.org/doi/10.1103/PhysRevLett.76.78}

\bibitem{Ferrenberg1}
A.~M. Ferrenberg, R.~H. Swendsen,
  \href{http://link.aps.org/doi/10.1103/PhysRevLett.61.2635}{New monte carlo
  technique for studying phase transitions}, Phys. Rev. Lett. 61 (1988)
  2635--2638.
\newblock \href {http://dx.doi.org/10.1103/PhysRevLett.61.2635}
  {\path{doi:10.1103/PhysRevLett.61.2635}}.
\newline\urlprefix\url{http://link.aps.org/doi/10.1103/PhysRevLett.61.2635}

\bibitem{Ferrenberg2}
A.~M. Ferrenberg, R.~H. Swendsen,
  \href{http://link.aps.org/doi/10.1103/PhysRevLett.63.1195}{Optimized monte
  carlo data analysis}, Phys. Rev. Lett. 63 (1989) 1195--1198.
\newblock \href {http://dx.doi.org/10.1103/PhysRevLett.63.1195}
  {\path{doi:10.1103/PhysRevLett.63.1195}}.
\newline\urlprefix\url{http://link.aps.org/doi/10.1103/PhysRevLett.63.1195}

\bibitem{privman1990}
V.~Privman, Finite Size Scaling and Numerical Simulation of Statistical
  Systems, World Scientific, 1990.

\bibitem{Itzykson83}
C.~Itzykson, R.~B. Pearson, J.~B. Zuber,
  \href{http://www.sciencedirect.com/science/article/pii/0550321383904996}{Distribution
  of zeros in ising and gauge models}, Nuclear Physics B 220~(4) (1983) 415 --
  433.
\newblock \href
  {http://dx.doi.org/http://dx.doi.org/10.1016/0550-3213(83)90499-6}
  {\path{doi:http://dx.doi.org/10.1016/0550-3213(83)90499-6}}.
\newline\urlprefix\url{http://www.sciencedirect.com/science/article/pii/0550321383904996}

\bibitem{Lee-Kosterlitz}
J.~Lee, J.~M. Kosterlitz,
  \href{http://link.aps.org/doi/10.1103/PhysRevB.43.3265}{Finite-size scaling
  and monte carlo simulations of first-order phase transitions}, Phys. Rev. B
  43 (1991) 3265--3277.
\newblock \href {http://dx.doi.org/10.1103/PhysRevB.43.3265}
  {\path{doi:10.1103/PhysRevB.43.3265}}.
\newline\urlprefix\url{http://link.aps.org/doi/10.1103/PhysRevB.43.3265}

\bibitem{Polymer13}
J.~C.~S. Rocha, S.~Schnabel, D.~P. Landau, M.~Bachmann,
  \href{http://link.aps.org/doi/10.1103/PhysRevE.90.022601}{Identifying
  transitions in finite systems by means of partition function zeros and
  microcanonical inflection-point analysis: A comparison for elastic flexible
  polymers}, Phys. Rev. E 90 (2014) 022601.
\newblock \href {http://dx.doi.org/10.1103/PhysRevE.90.022601}
  {\path{doi:10.1103/PhysRevE.90.022601}}.
\newline\urlprefix\url{http://link.aps.org/doi/10.1103/PhysRevE.90.022601}

\bibitem{Kenna_2}
R.~Kenna, A.~C. Irving,
  \href{http://www.sciencedirect.com/science/article/pii/037026939500316D}{Logarithmic
  corrections to scaling in the two dimensional xy-model}, Physics Letters B
  351~(1–3) (1995) 273 -- 278.
\newblock \href
  {http://dx.doi.org/http://dx.doi.org/10.1016/0370-2693(95)00316-D}
  {\path{doi:http://dx.doi.org/10.1016/0370-2693(95)00316-D}}.
\newline\urlprefix\url{http://www.sciencedirect.com/science/article/pii/037026939500316D}

\end{thebibliography}

\end{document}